\documentclass[pre,aps,showpacs,preprint]{revtex4-1}

\usepackage{fullpage}

\usepackage{graphicx}
\usepackage{amsmath}
\usepackage{amsfonts}
\usepackage{amssymb}
\usepackage{color}

\newcommand{\bea}{\begin{eqnarray}}
\newcommand{\eea}{\end{eqnarray}}
\newcommand{\beq}{\begin{equation}}
\newcommand{\eeq}{\end{equation}}
\newcommand{\bit}{\begin{itemize}}
\newcommand{\eit}{\end{itemize}}
\newcommand{\Y}{\mathcal{Y}}

\newcommand{\D}{\mathrm{d}}
\newcommand{\A}{\mathcal{A}}

\renewcommand{\P}{\mathcal{P}}
\renewcommand{\c}{\bar{c}}
\newcommand{\T}{\mathcal{T}}
\newcommand{\J}{\mathcal{J}}
\renewcommand{\r}{\right \rangle}
\renewcommand{\l}{\left \langle}

\begin{document}

\title{Fluctuation theorems and inequalities generalizing the second law of thermodynamics off equilibrium}

\author{Gatien Verley$^1$, David Lacoste$^1$}
\date{\today}

\begin{abstract}

We present a general framework for systems which are prepared in a non-stationary non-equilibrium state in the absence of any perturbation, and which are then further driven through the application of a time-dependent perturbation. We distinguish two different situations depending on the way the non-equilibrium state is prepared, either it is created by some driving; or it results from a relaxation following some initial non-stationary conditions. Our approach is based on a recent generalization of the Hatano-Sasa relation for non-stationary probability distributions.

We also investigate whether a form of second law holds for separate parts of the entropy production, in a way similar to the work of M. Esposito et al., Phys. Rev. Lett., 104:090601 (2010), but for a non-stationary reference process instead of a stationary one.
We find that although the special structure of the theorems derived in this reference is not recovered in the general case, detailed fluctuation theorems still hold separately for parts of the entropy production in this case. These detailed fluctuation theorems lead to interesting generalizations of the second-law of thermodynamics off equilibrium.

\end{abstract}

\maketitle

\section{Introduction}
In recent years, a broad number of works summarized under the name of fluctuations theorems, have lead  to significant progress in our understanding of the second law of thermodynamics  \cite{Jarzynski2011_vol2,Seifert2012_vol,Harris2007_vol2007}.
A central idea, namely the application of thermodynamics at the level of trajectories, has developed into a field of its own, called stochastic thermodynamics \cite{Seifert2012_vol,Seifert2008_vol64}.

In a similar spirit as the Crooks relation \cite{Crooks2000_vol61}, the total entropy production  can be expressed as the relative entropy between the probability distributions of trajectories associated with a forward and backward experiment \cite{Maes2003_vol110,Gaspard2004_vol117a,Kawai2007_vol98}. As a consequence,
the entropy production quantifies the time-symmetry breaking and
reversibility which means zero entropy production, only occurs when the forward and backward experiments are undistinguishable. While this statement for the entropy production encompasses the second law after averaging over many trajectories, it also provides additional implications at the trajectory level.


This particular idea has also played a central role in recent developments of the framework of fluctuation relations for
 systems operating under feedback control \cite{Sagawa2010_vol104}. A generalization of the Jarzynski relation \cite{Jarzynski1997_vol78} including the transfer of information due to feedback predicted theoretically in this reference has been tested experimentally \cite{Toyabe2010_vol6}. With these concepts, it is possible to reinterpret Landauer's principle linking information and thermodynamics \cite{Esposito2011_vol95}, and devise new experiments to test it in a particularly elegant and direct way \cite{Berut2012_vol483}.
Besides providing new insights into the deep connection between thermodynamics and information, progresses in stochastic thermodynamics make it possible to address optimization problems which should be relevant for many applications \cite{Aurell2011_vol106}.

In previous work, we have analyzed some consequences of a generalized Hatano-Sasa relation, in which the stationary distribution entering the original Hatano-Sasa relation \cite{Hatano2001_vol86} is replaced by a non-stationary one. In Ref.~\cite{Verley2011_vol}, we have shown that this relation offers a way to construct a modified fluctuation-dissipation theorem valid near an arbitrary non-equilibrium state; and in \cite{Verley2012_vol108}, we have also derived from it an interesting generalization of the second law of thermodynamics for non-stationary states. Such generalizations of the second law of thermodynamics and of the fluctuation-dissipation theorem are useful to describe the following situations: (i) the system is driven by at least two control parameters, so even when the driving of interest $h$ is constant in time, the probability distribution remains non-stationary and (ii) the system undergoes a transient regime due to the choice of initial conditions, and before the relaxation of this transient regime is finished, the system is further driven. We note that the second situation is typical of systems with a slow relaxation time, such as aging systems, in which case the system never reaches a stationary state on any reasonable time. Therefore, it seems to us that this framework should be ideally suited to analyze aging systems.

In this paper, we provide a more detailed analysis of the results of Ref.~\cite{Verley2012_vol108}, and we add some new applications. The first section contains preliminaries on fluctuation theorems. We then discuss a particular point concerning the symmetry property that the initial and final probability distributions should have for a detailed fluctuation theorem for the total entropy production to hold. Although this particular point is known in the literature \cite{Harris2007_vol2007}, it has been overlooked in many other works in the field despite its importance, and for this reason it seems to us that it was useful to provided a refreshing view about this somewhat subtle point.
In the next section, we discuss extensions of the non-adiabatic and adiabatic entropy productions which were introduced in Ref.~\cite{Esposito2010_vol104} for the case of a stationary reference process. We find that the special structure of the "three theorems" derived in this reference is not recovered in the general case of a non-stationary reference process. We interpret this as being due to the contribution of a new time-symmetric contribution in the dynamical action, which takes a form similar to the traffic introduced in Ref.~\cite{Maes2006_vol96}. We then discuss the second-law like inequalities which follow from the integral fluctuation theorems and which should be applicable to a broad class of non-equilibrium systems. In the last section, we present some illustrative examples of these ideas, using a two state model or a particle in an harmonic potential submitted to Langevin dynamics.

\section{Fluctuation theorems from general considerations of time-reversal symmetry}
\label{recipe}

\subsection{Stochastic modelling and definitions}
We consider a system which is assumed to evolve according to a continuous-time Markovian dynamics of a pure jump type \cite{Feller1940_vol48}. Let us introduce the transition rate $w_t(c,c')$ for the rate to jump from a state $c$ to a state $c'$ at time $t$. The subscript $t$ in $w_t(c,c')$ indicates that there are processes which are non-stationary even in the absence of explicit driving. The origin of such processes is arbitrary, they can result from an additional underlying driving, which is different from explicit driving and does not need to be specified. Note that if the system is submitted to an initial quench and a constant driving (explicit or not), the rates are time independent but evolution is still non stationary due to the initial quench.
At time $t=0$, an arbitrary explicit driving protocol $h_{t}$ is applied to the system, and we denote by $p_t(c,[h_t])$ the probability to observe the system in the state $c$ at a time $t$ in the presence of this driving. The evolution of the system for $t>0$ is controlled by the generator $L_t^{h_t}$, which is defined by
\beq
L^{h_t}_t(c',c) = w^{h_t}_t(c',c) - \delta(c,c')\sum_{c''} w^{h_t}_t(c',c''), \label{def L}
\eeq
where $w^{h_t}_t(c',c)$ is a transition rate in the presence of the driving $[h_t]$. Then $p_t(c,[h_t])$ is the solution of
\beq
\frac{d p_t(c,[h_t])}{dt} = \sum_{c'}  p_t(c',[h_t])  L^{h_t}_t(c',c).
\label{ME}
\eeq
The notation $p_t(c,[h_t])$ emphasizes that this probability distribution depends functionally on the whole protocol history $[h_t]$ up to time $t$. We assume that at $t=0$ there is no driving, so that $p_0(c,[h_0])=p_0(c)$. We also note that in practice, the driving $[h_t]$ may not start immediately at $t=0^+$ but may be turned on only later, after a certain time, called the waiting time in the context of aging systems.


We now introduce a different probability distribution denoted $\pi_t(c,h)$
which represents the probability to observe the system in the state $c$ at a time $t>0$ in the presence of a constant (time independent) driving $h$.
In other words, $\pi_t(c,h)$ follows from $p_t(c,[h_t])$ by freezing the time dependence in the driving $[h_t]$. This distribution, which will play a key role in the following, obeys the master equation
\beq
\left( \frac{\partial \pi_t}{\partial t}  \right) (c,h) = \sum_{c'}  \pi_t(c',h) L_t^{h}(c',c).
\label{defpi}
\eeq
From the fact that $\pi_t(c,h)$ and $p_t(c,[h_t])$ should coincide for a constant protocol, we deduce the initial condition to be $\pi_0(c,h_0) = p_0(c)$.

In Ref.~\cite{Verley2011_vol}, we have shown that one can construct with this distribution
the following functional
\beq
\Y[c] = - \int_0^T\D \tau \dot h_\tau \partial_h \ln \pi_\tau(c_\tau,h_\tau),
\label{defY}
\eeq
which has clear similarities with the functionals introduced by Jarzynksi \cite{Jarzynski1997_vol78} and Hatano-Sasa \cite{Hatano2001_vol86}.
We find from the analysis of this paper, that the functional $\Y$ has the interpretation of the driving part in the total entropy production.
Using a Feynman-Kac approach, which has also played a central role for the Jarzynski relation \cite{Hummer2001_vol98}, we have shown in ref.~\cite{Verley2011_vol} that this functional $\Y$ obeys a generalized Hatano-Sasa relation:
\beq
\langle \exp \left( -\Y [c] \right) \rangle =1
\label{modified HS}
\eeq

This relation qualifies for a generalization of the Hatano-Sasa relation because the stationary probability distribution $p_{st}(c,h)$ which enters in the functional $Y [c]= - \int_0^T \D t \, \dot h_t \partial_h \ln p_{st}(c_t,h_t)$ in the standard Hatano-Sasa relation is now replaced by the more general distribution $\pi_t(c,h)$. From a linear expansion of this generalized Hatano-Sasa, we have obtained modified fluctuation-dissipation theorems valid near an arbitrary non-equilibrium state \cite{Verley2011_vol,Chetrite2009_vol80}. In the next sections, we derive this generalized Hatano-Sasa relation in a different way and we investigate other consequences not contained in such a linear expansion.

\subsection{Path probability distributions and action functional}
\label{sec:path proba}

Let us consider a trajectory  $[c]=(c_0,c_1,...,c_N;\tau_1,.., \tau_N)$ where the $c_i$ are the states which are visited by the system and $\tau_i$ are the jumping times to go from $c_{i-1}$ to $c_i$. The total time-range of the trajectory is $[0..T]$. We denote $\P[c]$ the probability to observe such a trajectory $[c]$, also called path probability below :
\beq
\P[c]= p_0(c_0) \left[ \prod_{j=1}^{N} \exp \left( - \int_{\tau_{j-1}}^{\tau_{j}} \D \tau \lambda^{h_\tau}_\tau(c_{j-1}) \right) w^{h_{\tau_j}}_{\tau_j}(c_{j-1},c_{j}) \right]  \exp \left( - \int_{\tau_{N}}^{T} \D \tau \lambda_\tau^{h_\tau}(c_{N}) \right),
\label{path}
\eeq
where $\lambda^{h_\tau}_\tau(c')=\sum_{c \neq c'} w^{h_\tau}_\tau(c',c)$ represents the escape rate to leave the state $c'$, and $p_0(c_0)=p_0(c_0,h_0)$ represents the probability distribution of the initial condition.

In the following, we consider several ratios of path probabilities of the form:
\beq
\Delta \A[c]= \ln \frac{\P[c]}{\tilde{\P}[c^*]},
\label{def A}
\eeq
where the tilde symbol $(\sim)$ corresponds to a transformation of the original dynamics into a new dynamics. This new dynamics is defined by its own initial condition and by the transformed transition rates denoted by $\tilde{w}$. The $(*)$ denotes a different transformation which acts on the trajectory itself. The transformed trajectory $[c^*]=(c_0^*,c_1^*,..,c_N^*;\tau_1^*,.., \tau_N^*)$ results from the application of an involution on the trajectory $[c]$ which we assume to be either the identity ($[c^*]=[c]$) or the time-reversal symmetry acting on the trajectories ($[c^*]=[\bar{c}]=(c_N,c_{N-1},..,c_0;T-\tau_N,..,T-\tau_1)$). In other words, we have
\beq
c^*_i = \left\lbrace
\begin{array}{ll}
c_i & \mbox{if $*$ is identity,} \\
c_{N-i} \quad & \mbox{if $*$ is time reversal,}
\end{array} \right. \qquad
\tau^*_i = \left\lbrace
\begin{array}{ll}
 \tau_i & \mbox{if $*$ is identity,} \\
\tau_{N-i+1} \quad & \mbox{if $*$ is time reversal,}
\end{array} \right.
\eeq
with the convention that $\tau^*_{0}$ and $\tau^*_{N+1}$ are respectively $0$ and $T$ when $*$ is identity and are respectively $T$ and $0$ when $*$ is the time reversal symmetry.
Substituting the trajectory $[c^*]$ in replacement of $[c]$, and the rates of the modified dynamics $\tilde w$ instead of the original rates $w$ in Eq.~\ref{path}, one obtains directly for the transformed path probability :
\beq
\tilde{\P}[c^*]= \tilde{p}_0(c_0^*) \left[ \prod_{j=1}^{N} \exp \left( - \int_{\tau^*_{j-1}}^{\tau^*_{j}} \D \tau \tilde{\lambda}^{h_\tau}_{\tau}(c_{j-1}^*) \right) \tilde{w}^{h_{\tau_j^*}}_{\tau_j^*}(c_{j-1}^*,c_{j}^*) \right]  \exp \left( - \int_{\tau^*_{N}}^{\tau^*_{N+1}} \D \tau \tilde{\lambda}_{\tau}^{h_\tau}(c_{N}^*) \right),
\eeq
where $\tilde \lambda^{h_\tau}_\tau(c')=\sum_{c \neq c'} \tilde w^{h_\tau}_\tau(c',c)$ represents the escape rate to leave the state $c'$ in the dynamics modified via the operation tilde.
From this we see that $\Delta \A[c]$ can be written as
\beq
\Delta \A[c] = \ln \frac{p_0(c_0)}{\tilde{p_0}(c_0^*)} - \int_0^T dt [\lambda^{h_t}_t(c_t) - \tilde{\overset{*}{\lambda}}\,^{h_t}_t(c_t) ] + \sum_{j=1}^N \ln \frac{w_{\tau_j}^{h_{\tau_j}}(c_{j-1},c_j)}{\tilde{w}_{\tau_j^*}^{h_{\tau_j^*}}(c_{j-1}^*,c_j^*)},
\label{explicit A}
\eeq
with $c_t=c_j$ if $t \in [\tau_j,\tau_{j+1}[$ and $ \overset{*}{\lambda}\,\!^{h_t}_t = \lambda^{h_{T-t}}_{T-t}$ (or $ \overset{*}{\lambda}\,\!^{h_t}_t = \lambda^{h_{t}}_{t}$) if the involution $*$ is the time reversal (or respectively if $*$ is identity). Thus, $\Delta \A[c]$ has three different contributions: the first term is a boundary term which only depends on the initial or final configurations, the last term is a bulk term, which depends on the whole trajectory. The second term is related to the notion of traffic \cite{Baiesi2009_vol103}, which represents the integral of the escape rate $\lambda_t$ evaluated at the actual configuration $c_t$ of the system at time $t$. In view of this property, the second term in Eq.~\ref{explicit A} represents a difference of traffic between the original dynamics (which corresponds to $\P$) and the transformed dynamics (which corresponds to $\tilde{\P}$).

\subsection{Protocol-reversal symmetry and the probability distributions of the initial and final points}

Fluctuations theorems can be derived from considerations of symmetry for an arbitrary observable and arbitrary initial and final probability distributions \cite{Seifert2005_vol95}. These choices of observables, of initial and final probability distributions determine precisely which fluctuation theorem holds. In this construction, we emphasize that the fluctuation theorem takes a strong form if the initial and final probability distributions are related by a reversal of protocol and a weaker form if not \cite{Harris2007_vol2007}. Then, two cases must be considered, either the initial and final path probabilities are not related by the reversal of the protocol and the transformation ($\sim$) is not an involution; or such a symmetry exists and the transformation is an involution. In the following, we discuss both cases separately :

\begin{itemize}
  \item Let us first assume that $\sim$ is not an involution. This occurs for instance when the initial condition does not satisfy $\tilde{\tilde p}_0(c) = p_0(c)$. Following Ref.~\cite{Esposito2010_vol104}, we consider
\bea
P(\Delta \A[c] = \Delta A) &=& \sum_{[c]} \delta(\Delta \A - \Delta \A[c]) \P[c], \\
&=& \exp \left( \Delta \A \right) \sum_{[c]} \delta(\Delta \A - \Delta \A[c] )  \tilde{\P}[c^*], \\
&=&  \exp \left( \Delta \A \right) \tilde{P}(\Delta \A[c^*] = \Delta \A),
\label{DFTNI}
\eea
with
\beq
\tilde{P}(\Delta \A[c^*] = \Delta \A) =\sum_{[c]} \delta(\Delta \A - \Delta \A[c^*] )  \tilde{\P}[c].
\label{Ptilde2}
\eeq
With words, $\tilde{P}(\Delta \A[c^*] = \Delta \A)$ corresponds to the probability to have  on a given trajectory $[c^*]$, $\Delta \A[c^*]$ equal to $\Delta \A$ in the tilde experiment/dynamics. When comparing with the expression of $P(\Delta \A[c] = \Delta A)$, it appears that the same function $\Delta \A$ is evaluated on different trajectories ($[c]$ or $[c^*]$), which are themselves generated by different dynamics (the original dynamics or the tilde dynamics).
Thus, the probability $\tilde P(\Delta \A[c^*] = \Delta \A) $ cannot be defined in itself, i.e. without reference to the quantity $\Delta \A$ introduced in the original dynamics \cite{Harris2007_vol2007}. For this reason, we regard the detailed fluctuation theorem (DFT) of Eq.~\ref{DFTNI} has a weak version of the theorem.

  \item Let us then assume that the operation ($\sim$) is an involution acting on the path probabilities, $\tilde{\tilde \P} = \P$. This implies that the distribution of initial condition satisfies the condition $\tilde{\tilde p}_0(c) = p_0(c)$ and that the transition rates satisfy $\tilde{\tilde{w}}^{h_t}_t(c,c')=w^{h_t}_t(c,c')$. From these two conditions or equivalently directly from the definition Eq.~\ref{def A}, it follows that:
\beq
\Delta \A[c]=-\Delta \tilde{\A}[c^*],
\label{sym_property}
\eeq
where $\Delta \tilde{\A}[c] = \ln \tilde{\P}[c]/\tilde{\tilde{\P}}[c^*] $.
With this symmetry property, the fluctuation relation for $\Delta \A$ now takes the form
\bea
P(\Delta \A[c] = \Delta A) &=&  \exp \left( \Delta \A \right) \sum_{[c]} \delta(\Delta \A + \Delta \tilde \A[c^*] )  \tilde{\P}[c^*], \\
&=&  \exp \left( \Delta \A \right) \tilde{P}(\Delta \tilde{\A}[c] = -\Delta \A),
\label{DFTI}
\eea
with
\beq
\tilde{P}(\Delta \tilde{\A}[c] = -\Delta \A) =\sum_{[c]} \delta(\Delta \A + \Delta \tilde \A[c] )  \tilde{\P}[c],
\label{Ptilde1}
\eeq
which corresponds with words to the probability to have on a given trajectory $[c]$, $\Delta \tilde{\A}[c]$ equal to $ -\Delta \A$ in the tilde experiment/dynamics.
As expected one can obtain directly Eq.~\ref{Ptilde1} from Eq.~\ref{Ptilde2} using Eq.~\ref{sym_property}.
The main difference with the previous case where tilde was not an involution is that now, it is not the same function which must be evaluated in the two experiments/dynamics characterized by $P$ (resp. $\tilde{P}$); rather it is two different functions, namely $\Delta \A[c]$ and $\Delta \tilde{\A}[c]$) but they are related because they represent  the same physical quantity which takes different form on each experiment/dynamics.
This is similar to the Crooks relation \cite{Crooks2000_vol61,Horowitz2007_vol}, where the same physical concept, namely the dissipated work, must be evaluated in the direct and tilde experiment/dynamics (although the precise function which represents this physical concept takes a different form in both cases). The main point is that here unlike in the previous case, the function which must be evaluated is linked to the process (direct or reversed) under consideration. We thus regard Eq.~\ref{DFTI} as a strong form of the detailed fluctuation theorem.
\end{itemize}

As a particular important illustration of this point, we discuss below the detailed fluctuation theorem satisfied by the entropy production.
To do so, we consider both involutions introduced above, namely $(*)$ and $(\sim)$, to represent a reversal symmetry, respectively the reversal of trajectories and of protocol, which we both denote with a bar $(-)$. We recall that the effect of this symmetry must be considered separately on the trajectories and on the dynamics. The rates which control the dynamics are transformed as
\beq
\bar{w}^{h_\tau}_\tau(c,c') = w^{h_{T-\tau}}_{T-\tau}(c,c'),
\label{time reversed rates}
\eeq
since the order in the visited configurations is not affected by the transformation while the time dependance of the driving is. Therefore, one can think of this transformation as basically a time-reversal of all protocols (the driving $[h_t]$ and the other protocols represented by the extra subscript in the rates). Note also that Eq.~\ref{time reversed rates} represents a transformation for the rates which is always an involution unlike the full reversal of the path probabilities which may or may not be an involution depending on the initial conditions.
This point is very relevant for the existence of a detailed fluctuation theorem for the entropy production. Indeed, in order to identify $\A$ as entropy production, the initial probability distribution of the reversed process must correspond to the final probability distribution reached by the direct process \cite{Seifert2005_vol95}. In other words, one must choose $\bar{p_0}(\bar{c_0})=p_T(c_T)$ where $p_T$ is the solution of the Master equation/Fokker Planck equation at time $T$.
From Eq.~\ref{explicit A}, due to the vanishing of the second term, one obtains the familiar result
\cite{Maes2003_vol110,Seifert2005_vol95}:
\beq
\Delta S_{tot}[c]= \ln \frac{\P[c]}{\bar{\P}[\c]}= \Delta S
+ \sum_{j=1}^N \ln \frac{w_{\tau_j}^{h_{\tau_j}}(c_{j-1},c_j)}
{w_{\tau_j}^{h_{\tau_j}} (c_j,c_{j-1})},
\label{tep}
\eeq
where the first term $\Delta S=\ln {p_0(c_0)} - \ln {p_T(c_T)}$ represents the change in system  stochastic entropy while the second term represents the change in reservoir entropy $\Delta S_r[c]$ along the specified trajectory [c].

In view of the discussion above, it is not obvious that the transformation of the full path probability denoted $(-)$ as defined above is an involution in the particular case of the entropy production.
Only when additional assumptions are made, namely that the initial and final probability distributions are related by a reversal of the protocol, can this transformation be an involution. Incidentally, this condition means equivalently that the system stochastic entropy $\Delta S$ is antisymmetric with respect to a reversal of the protocol. When this is the case, one obtains from Eq.~\ref{DFTI}, the following detailed fluctuation relation
\beq
\ln \frac{P(\Delta S_{tot}[c]=\Delta S_{tot})}{\bar{P} (\Delta \bar{ S}_{tot}[c] = -\Delta S_{tot})}= \Delta S_{tot},
\label{DFT Stot}
\eeq
which many authors as \cite{Esposito2010_vol104} have denoted using a simplified notation
\beq
\ln \frac{P(\Delta S_{tot})}{\bar{P} ( -\Delta S_{tot})}= \Delta S_{tot}.
\label{DFT Stot2}
\eeq
Note that this relation takes the form of the Evans and Searles theorem \cite{Evans2002_vol51} in the following particular cases: (i) for non-equilibrium stationary processes and (ii) for processes generated by time-symmetric driving protocols with the additional condition that the initial and final conditions are related by the reversal of the protocol. 

When $p_0(c_0)$ and $\bar{p_0}(\bar{c_0})$ are not related by a protocol reversal, the detailed fluctuation theorem for the entropy production only holds in its weak form namely Eq.~\ref{DFTNI}. As explained above, this means that the quantity which enters this detailed fluctuation theorem for the reversed process is not the entropy production of that process.

\subsection{Dual dynamics and difference of traffic}
\label{dual dynamics}

We now introduce a new transformation, called a duality transformation, which acts specifically on the dynamics of the process. In the following, this transformation is denoted with a hat ($\wedge $). In analogy with the way this dual transformation has been introduced in the stationary case \cite{Esposito2010_vol104,Hatano2001_vol86}, we define the dual dynamics
from the original dynamics by substituting the original rates $w^{h}_\tau(c,c')$ by :
\beq
\hat w^{h}_\tau(c,c')= \frac{w^h_\tau(c',c)\pi_\tau(c',h)}{\pi_\tau(c,h)}.
\label{duality}
\eeq
From this definition, it is not obvious that the duality transformation is an involution although it is indeed the case as we show in appendix \ref{dual current}. The basic idea is that this transformation essentially reverses the probability currents defined with respect to $\pi_t(c,h)$, and because of this, it follows that this transformation is an involution. The proof also confirms that the dynamics constructed from the dual rates is Markovian. The generator of the dynamics still verify $\sum_{c'} \hat{L}_t^{h_t}(c,c') = 0$, where we have defined $\hat L_t^{h_t}$ as in Eq.~\ref{def L} substituting the rates $w_t^{h_t}$ by the rates $\hat w_t^{h_t}$. The normalisation of the probability distribution is thus conserved in time.

An important property of the probability distribution $\pi_t(c,h)$ justifying its use to define the duality transform, is that it is related to the difference between the escape rates of the direct and dual dynamics, because :
\bea
\hat \lambda^{h}_{\tau}(c)- \lambda^{h}_{\tau}(c) &=& \sum_{c' \neq c} \left( \hat w^{h}_\tau(c,c') - w^{h}_\tau(c,c') \right), \\
&=& \sum_{c' \neq c} \left(\pi^{-1}_\tau(c,h) w_\tau (c',c) \pi_\tau(c',h) - w^{h}_\tau(c,c') \right), \\
&=& \sum_{c'} \pi^{-1}_\tau(c,h) w_\tau (c',c) \pi_\tau(c',h) - \sum_{c''} w^{h}_\tau(c,c''),\\
&=& \sum_{c'} \pi^{-1}_\tau(c,h) \left( w_\tau (c',c) - \delta_{c'c}\sum_{c''} w^{h}_\tau(c',c'') \right) \pi_\tau(c',h), \\
&=& \pi^{-1}_\tau(c,h) \left( \partial_\tau \pi_\tau \right) (c,h)= \left(\partial_\tau \ln \pi_\tau \right)(c,h), \label{traffic_pi_def}
\eea
where in the last step we used the evolution equation Eq.~\ref{defpi}. We define the difference of traffic between the direct and dual dynamics as
\beq
\Delta \T[c] =  \int_0^T \D \tau \left( \lambda^{h_\tau}_{\tau}(c_\tau) - \hat \lambda^{h_\tau}_{\tau}(c_\tau) \right) = -\int_{0}^{T} \D \tau \left(\partial_\tau \ln \pi_\tau \right)(c_\tau,h_\tau).
\label{explicit T}
\eeq
From this last expression, we note that the difference of traffic vanishes when the reference probability $\pi_t$ is stationary; and that this quantity is \emph{antisymmetric} under the duality transformation $\Delta \hat{\T}[ c] = - \Delta \T[c]$ but \emph{symmetric} under the combined action of the reversal of the trajectories and of the full protocol (which we regard as the  total-reversal symmetry):
\bea
\Delta \bar \T [\bar c]&=&  \int_0^T d\tau \left( \bar \lambda_{\tau}^{ h_{\tau}}(\bar c_{\tau}) - \hat{\bar \lambda}_{\tau}^{h_{\tau}}(\bar c_{\tau})  \right), \\
&=& \int_0^T d\tau \left( \lambda_{T-\tau}^{ h_{T-\tau}}(c_{T-\tau}) - \hat \lambda_{T-\tau}^{h_{T-\tau}}(c_{T-\tau})  \right), \\
&=& \Delta \T [c], \label{sym_traffic}
\eea
In the end, combining the total-reversal symmetry and the duality transform together, we obtain that $ \Delta \hat{\bar \T} [\bar c] = - \Delta \T [c]$.

\subsection{Adiabatic and non-adiabatic entropy productions}
A system can fall into a non-equilibrium state by two mechanisms: (i) either detailed balance can be broken due for instance to boundary conditions or (ii) the system can be driven. Building on a number of works on steady-state thermodynamics \cite{Hatano2001_vol86,Oono1998_vol,Speck2005_vol38,Chernyak2006_vola}, it was shown in Ref.~\cite{Esposito2010_vol104} that that these two different ways to put a system in a non-equilibrium state correspond to two separate contributions in the entropy production, called adiabatic for case (i) and non-adiabatic for case (ii). Note that this term "adiabatic" does not refer to the absence of heat exchange but rather to the fact that this contribution is the only one which remains in the adiabatic limit of very slow driving. In this reference, it was shown that surprisingly both terms can be expressed as logratios of probabilities, which implies that both quantities satisfy separately a detailed fluctuation theorem (DFT). This property is surprising because it is not expected to hold for a general splitting of the entropy production. Indeed it does not hold for instance for the splitting of the entropy production into system entropy and reservoir entropy \cite{Seifert2005_vol95}. As a further consequence of these DFTs, both the adiabatic part and the non-adiabatic are positive on average, which means that the second law can be split into these two components.

In this section, we generalize the notions of adiabatic and non-adiabatic entropy productions defined as in \cite{Esposito2010_vol104,Esposito2007_vol76} for the stationary case, by replacing the stationary distribution by the distribution $\pi_t(c,h)$ defined in Eq.~\ref{defpi}. We obtain the following splitting
\beq
\Delta S_{na}[c] =  \ln \frac{p_0(c_0)}{p_T(c_T,[h_T])}
+ \sum_{j=1}^N \ln \frac{\pi_{\tau_j}(c_j,h_{\tau_j})}
{\pi_{\tau_j}(c_{j-1},h_{\tau_j})},
\label{Sna}
\eeq
and
\beq
\Delta S_a[c] =  \sum_{j=1}^N \ln \frac{w_{\tau_j}^{h_{\tau_j}}(c_{j-1},c_j) \pi_{\tau_j}(c_{j-1},h_{\tau_j})}
{w_{\tau_j}^{h_{\tau_j}} (c_j,c_{j-1}) \pi_{\tau_j}(c_j,h_{\tau_j})},
\label{Sa}
\eeq
so that we still have $\Delta S_{tot}[c]=\Delta S_{na}[c]+ \Delta S_{a}[c]$.

We note that the adiabatic entropy production verify $\Delta \bar{S}_{a}[\bar{c}]=- \Delta S_{a}[c]$ which means that it is anti-symmetric with respect to the combination of the protocol-reversal and the time-reversal of the trajectories, transformation that we call the total-reversal. On the other side, the non-adiabatic entropy production is anti-symmetric under the total-reversal, i.e.  $\Delta \bar{S}_{na}[\bar{c}]=- \Delta S_{na}[c]$,
when the total entropy is (provided the appropriate condition on the initial and final states holds as explained in the previous section). We can also define an excess entropy production $\Delta S_{ex}$ such that $\Delta S_{na}[c] = \Delta S + \Delta S_{ex}[c]$ and $\Delta S_{a}[c] = \Delta S_r[c] - \Delta S_{ex}[c]$.


It is natural to ask at this point whether $\Delta S_{na}$ and $\Delta S_{a}$ separately satisfy a DFT. These quantities are not a priori of the form of Eq.~\ref{def A}, except for the particular case studied in \cite{Esposito2010_vol104} where the reference is stationary, so $\Delta S_{na}$ and $\Delta S_{a}$ should thus not in general satisfy separately a DFT. We thus loose, with the definition of Eqs.~\ref{Sna}-\ref{Sa}, the positivity of the mean adiabatic and non adiabatic entropy productions.
Despite this, we will see that their joint probability distribution still satisfies a DFT as explained in section \ref{jointFT}.

\subsection{Non-adiabatic and adiabatic action functionals}

\label{action functionals}

In this section, we show that the difference of traffic $\Delta \T$ introduced above is a key observable which can be used to construct quantities which satisfy a DFT. We start from the two possible decompositions of the entropy production as
\beq
\mbox{(A) } \quad \Delta S_{tot}[c] = \ln \frac{\P[c]}{\hat{\bar \P}[\bar c]} + \ln \frac{\hat{\bar \P}[ \bar c]}{\bar \P[\bar c]} \qquad \mbox{ or \; (B) } \qquad  \Delta S_{tot}[c] = \ln \frac{\P[c]}{\hat{\P}[c]} + \ln \frac{\hat{\P}[c]}{\bar \P[\bar c]}.\label{choice}
\eeq
We first remark that, contrary to the case of \cite{Esposito2010_vol104} where the stationary probability distribution is chosen as a reference, the two decompositions are not equivalent. This is due to the fact that the two terms in the decomposition are not anti-symmetric under total-reversal any more, since they contain a non zero difference of traffic term, defined above:
$$\ln \frac{\hat{\bar \P}[ \bar c]}{\bar \P[\bar c]} \neq \ln \frac{\P[c]}{\hat{\P}[c]} \qquad \mbox{and} \qquad \ln \frac{\hat{\P}[c]}{\bar \P[\bar c]} \neq \ln \frac{\P[c]}{\hat{\bar \P}[\bar c]}.  $$

\textbf{Case A} -
We first focus on the first term in the r.h.s. of Eq.~\ref{choice}A, which we call the non-adiabatic action $\Delta A_{na}[c]$. Using Eq.~\ref{explicit A} with the choice $\tilde{\P}=\hat{\bar{\P}}$ for the path probabilities and $c^*=\bar{c}$ for the trajectories, we obtain
\beq
\Delta A_{na}[c]=\ln \frac{\P[c]}{\hat{\bar \P}[\bar c]} = \ln \frac{p_0(c_0)}{\hat{\bar p}_0(\bar{c_0})} - \int_{0}^{T} \D \tau \left( \lambda^{h_\tau}_\tau(c_{\tau}) - \hat \lambda^{h_\tau}_\tau(c_{\tau})\right) +  \sum_{j=1}^{N} \ln \frac{w^{h_{\tau_j}}_{\tau_j}(c_{j-1},c_{j})}{\hat w^{h_{\tau_j}}_{\tau_j}(c_{j},c_{j-1}) }.
\label{def Ana}
\eeq
Given the initial condition $\hat{\bar p}_0(\bar{c_0})=p_T(c_T)$ for the dual reversed experiment, the first term in this equation corresponds to
what we have denoted before $\Delta S=\ln p_0(c_0)- \ln p_T(c_T)$. Using Eq.~\ref{duality} and Eq.~\ref{Sna}, we obtain
\bea
\Delta A_{na}[c] &=& \Delta S + \int_{0}^{T} \D \tau \left(\partial_\tau \ln \pi_\tau \right)(c_\tau,h_\tau) + \sum_{j=1}^{N} \ln \frac{ \pi_{\tau_j}(c_{j},h_{\tau_j})}{ \pi_{\tau_j}(c_{j-1},h_{\tau_j})}, \nonumber  \\
&=& \Delta S_{na}[c] - \Delta \T[c],   \label{nadb decomp}
\eea
which corresponds to a decomposition into two terms, where the first term, $\Delta S_{na}[c]$, is anti-symmetric and the second term, $\Delta \T[c]$ symmetric under total-reversal.
Alternatively, we can also write the same quantity as
\bea
\Delta A_{na}[c] &=& \Delta S + \int_{0}^{T} \D \tau \left(\partial_\tau \ln \pi_\tau \right)(c_\tau,h_\tau) + \sum_{j=1}^{N} \ln \frac{ \pi_{\tau_j}(c_{j},h_{\tau_j})}{ \pi_{\tau_j}(c_{j-1},h_{\tau_j})}, \label{simplifying}  \\
&=& \Delta S - \Delta \psi + \int_0^T d\tau \partial_\tau \left( \psi_\tau(c_\tau,h_\tau) \right) + \int_{0}^{T} \D \tau \left(\partial_\tau \ln \pi_\tau \right)(c_\tau,h_\tau) \nonumber \\
 &&  + \sum_{j=1}^{N} \ln \frac{ \pi_{\tau_j}(c_{j},h_{\tau_j})}{ \pi_{\tau_j}(c_{j-1},h_{\tau_j})} \nonumber \\
 &=& \Delta S_b + \Y[c], \label{naut_DFR}
\eea
where $\Delta S_b = \Delta S - \Delta \psi$ is a boundary term, with $\Delta \psi=- \ln \pi_T(c_T,h_T)+\ln \pi_0(c_0,h_0) $.

Therefore, since $p_0(c_0)=\pi_0(c_0,h_0)$ by construction,
\beq
\Delta S_b= \ln \frac{\pi_T(c_T,h_T)}{p_T(c_T)}.
\eeq
As a result, the average of $\Delta S_b$, is related to the Kullback-Leibler divergence between the distributions $\pi_T$ and $p_T$. Physically, this quantity can be viewed as a measure of the lag between the two distributions, in the same way that one can look at the dissipated work as a measure of the lag between the actual distribution at time $t$ and the corresponding equilibrium distribution with the control parameter at the same value \cite{Vaikuntanathan2009_vol87}. When there is no lag, either because $p_T$ has relaxed towards $\pi_T$, or because the initial probability distribution of the reversed protocol, namely $p_T(c_T)$, is chosen to be $\pi_T(c_T,h_T)$, then the two distributions are identical and $\Delta S_b$ vanishes. In this case, $\Delta A_{na}[c]=\Y[c]$, which satisfies the symmetry condition $\hat{\bar{\Y}}[\bar c]=-\Y[c]$. Therefore, from Eq.~\ref{def A} and Eq.~\ref{DFTI}, this quantity satisfies a DFT:
\beq
\ln \frac{P(\Y[c] = \Y)}{\hat{\bar{P}} ( \hat{\bar \Y}[c] = -\Y)}= \Y.
\label{DFT A_na}
\eeq
If we don't have a vanishing boundary term, then unfortunately only the weak fluctuation theorem of Eq.~\ref{DFTNI} is verified
\beq
\ln \frac{P(\Delta \A[c] = \Delta A_{na})}{\hat{\bar{P}} (\Delta \A[\bar c] = \Delta A_{na})}= \Delta A_{na}.
\label{DFT A_nabis}
\eeq


We now look at the second term in Eq.~\ref{choice}A, namely
\beq
\Delta A_{a}[c] = \Delta S_{tot}[c]-\Delta A_{na}[c]=\ln \frac{\hat{\bar \P}[\bar c]}{\bar \P[\bar c]}.
\label{aut_DFR}
\eeq
We can rewrite Eq.~\ref{aut_DFR} as
\bea
\Delta A_a[c] &=&  \int_{0}^{T} \D \tau (\lambda^{h_\tau}_\tau(c_{\tau}) - \hat \lambda^{h_\tau}_\tau(c_{\tau})) +  \sum_{j=1}^{N} \ln \frac{\hat{\bar{ w}}^{ h_{T-\tau_j}}_{T-\tau_j}(c_{j},c_{j-1})}{\bar{w}^{h_{T-\tau_j}}_{T-\tau_j}(c_{j},c_{j-1}) }, \\
&=& - \int_{0}^{T} \D \tau \left(\partial_\tau \ln \pi_\tau \right)(c_\tau,h_\tau) +  \sum_{j=1}^{N} \ln \frac{  w^{ \, h_{\tau_j}}_{\tau_j}(c_{j-1},c_{j}) \pi_{\tau_j}(c_{j-1},h_{\tau_j})}{w^{\, h_{\tau_j}}_{\tau_j}(c_{j},c_{j-1})  \pi_{\tau_j}(c_{j},h_{\tau_j}) }, \label{def_aut} \\
&= & \Delta S_a[c] + \Delta \T[c], \label{adb decomp}
\eea
which corresponds again to a decomposition where the first term, $\Delta S_a[c]$, is anti-symmetric and the second term, $\Delta \T[c]$ symmetric under total-reversal. As a self-consistent check, we see that the difference of traffic $\Delta \T$, in $\Delta A_a$ exactly compensates an opposite contribution in $\Delta A_{na}$ so that
\beq
\Delta S_{tot}=\Delta A_a + \Delta A_{na}=\Delta S_{na} + \Delta S_a.
\label{decomp1}
\eeq
One important point is to realize that $\Delta A_a$ is not of the form of Eq.~\ref{def A} because it involves a modified path probability both at the numerator and the denominator in its definition. Therefore a detailed fluctuation relation of the form of Eq.~\ref{DFT A_na} is \textit{not} verified for this quantity.

\vskip 0.5cm
\textbf{Case B} -
One can however find another DFT, by starting from the splitting of the entropy production of Eq.~\ref{choice}B. We define the first term on the r.h.s. by
\beq
\Delta B_a[c]  = \ln \frac{\P[c]}{\hat \P[c]} = - \Delta \bar{A}_{a} [\bar c] = - \Delta \T[c] +\Delta S_{a}[c].
\label{defBa}
\eeq
and we also introduce the quantity
\beq
\Delta B_{na}[c] = \ln \frac{\hat{\P}[c]}{\bar \P[\bar c]} = \Delta \hat{A}_{na}[c] = \Delta \T[c]+ \Delta S_{na}[c].
\eeq
Here $\Delta B_{a}$ plays a role similar to $\Delta A_{na}$ since it too has the required form to satisfy a detailed FT, which is
\beq
\ln \frac{P(\Delta B_{a}[c] = \Delta B_{a})}{\hat{P} (\Delta \hat B_{a}[c] = -\Delta B_{a})}= \Delta B_{a}.
\label{DFT B_a}
\eeq
As before for $\Delta A_{a}$, the remaining part in the total entropy, namely $\Delta B_{na}$, does not satisfy a detailed FT.

\subsection{Some limiting cases of interest}
In this section, we discuss some of the limiting cases for which the detailed fluctuation relations obtained above simplify.
Let us assume that the driving starts at time $t_{di}>0$ and ends at time $t_{df}<T$ for a total duration $t_d=t_{df}-t_{di}$.

\begin{itemize}
  \item When $\pi_t(c,h)$ relaxes very quickly to the stationary distribution (on a time scale $\tau_{st}$ such that $\tau_{st} \ll T$ and $\tau_{st} \ll t_d$), one recovers from Eq.~\ref{Sna} and Eq.~\ref{Sa} the usual definitions of the non-adiabatic and adiabatic parts of the entropy production. In this case $\Delta \T=0$, and as a result Eq.~\ref{DFT A_na} and Eq.~\ref{DFT B_a} become the usual DFTs satisfied by the non-adiabatic and adiabatic entropies respectively \cite{Esposito2010_vol104}.


  \item In the limit of slow driving $\dot h_t \simeq 0$, which can happen without having $\pi_t(c,h)$ relaxed to a stationary distribution, the driving part of the entropy production,
$\Y[c]$, vanishes. Furthermore, the boundary term $\Delta S_b$ also vanishes, because in this case $p_t(c,[h_t])$ relaxes to $\pi_t(c,h)$ since $[h_t] \rightarrow h$.
In this limit $\Delta A_{na}=0$, which justifies a posteriori the name non adiabatic action for $\Delta A_{na}$. The vanishing of $\Delta A_{na}$ has two further consequences, the first one is that Eq.~\ref{def Ana} implies $\P[c]=\hat{\bar{\P}}[\bar{c}]$, in other words, the duality and total-reversal compensate each other exactly. Another consequence is that $\Delta A_a=\Delta B_a=\Delta S_{tot}$, which implies that $\Delta \T=0$ although $\pi_t(c,h)$ is time-dependent. Furthermore, the fluctuation theorem of entropy production namely, Eq.~\ref{DFT Stot} coincides with that for $\Delta B_a$, namely Eq.~\ref{DFT B_a}.
\end{itemize}

\subsection{Modified second law for transition between non-stationary states}

The observables introduced in section \ref{action functionals} verify an integral Fluctuation Theorem and therefore are submitted to second law like inequalities, which are valid for an arbitrary non-equilibrium reference process. This results from the positivity of the Kullback-Leibler divergence between the path probabilities $\P[c]$ and $\hat{ \bar \P}[\bar c]$:
\beq
D(\P[c] || \hat{ \bar \P}[\bar c]) \equiv \sum_{[c]} \P[c] \ln \frac{\P[c]}{\hat{ \bar \P}[\bar c]} = \langle \Delta A_{na} \rangle  \ge 0,
\label{ineq0}
\eeq
and between the distributions $\P[c]$ and $\hat \P [c]$
\beq
D(\P[c] || \hat{ \P}[ c])=\sum_{[c]} \P[c] \ln \frac{\P[c]}{\hat \P[c]} = \langle \Delta B_a \rangle  \ge 0
\label{ineq2}
\eeq
In view of Eqs.~\ref{nadb decomp}, \ref{defBa} and \ref{decomp1} , this implies
\beq
\left\langle \Delta S_{na}\right\rangle   \ge \left\langle \Delta \T\right\rangle.
\label{ineq4}
\eeq
Furthermore, one also has $\left\langle \Delta S_{a} \right\rangle   \ge \left\langle \Delta \T\right\rangle$ and $ \left\langle \Delta S_{tot} \right\rangle \ge 0$, which taken together imply
$\left\langle \Delta S_{tot} \right\rangle \ge \max(2\langle \Delta \T \rangle,0).$
However, all these inequalities are less binding than Eq.~\ref{ineq4}, because the adiabatic entropy production and the total entropy production are generally increasing function of time whereas the inequality Eq.~\ref{ineq4} becomes an equality in the long time limit as explained below. We note furthermore that :

(i) Although we have shown that $\langle \Delta A_{na} \rangle  \ge 0$ and $\langle \Delta B_{a} \rangle  \ge 0$, the corresponding conjugate quantities of $\Delta A_{na}$ and $\Delta B_a$ with respect to the total entropy production, namely $\Delta A_a$ and $\Delta B_{na}$, do not have likewise a positive mean in general.
That this should be the case can be understood from the consideration of a particular case, namely the case where the initial non-equilibrium condition has been prepared by the application of a protocol, which is exactly compensated by the second protocol (the perturbation) denoted $[h_t]$ in this paper. In this case, the system is in equilibrium at all times in the presence of the perturbation. Since the rates satisfy the detailed balance condition, one can check explicitly that this implies $\Delta A_a=-\Delta A_{na}$ as expected since the system is in equilibrium and $\Delta S_{tot}=0$.
It follows from this that in this case $\langle \Delta A_{a} \rangle  \le 0$ and $\langle \Delta B_{na} \rangle  \le 0$; so in this case $\Delta A_a$ and $\Delta B_{na}$ do not have positive means.

(ii) The first inequality in Eq.~\ref{ineq4} can be written
\beq
\langle \Y \rangle \ge D(p_T||\pi_T) \ge 0.
\label{lag}
\eeq
In other words, the average of the functional $\Y$ is bounded by $-\Delta S_b=D(p_T||\pi_T)$ which is a measure of the lag between the distributions $p_T$ and $\pi_T$. As noted before, a similar result holds for the dissipated work for the case of an initial equilibrium probability distribution \cite{Vaikuntanathan2009_vol87}. Recalling the definition of the excess entropy,
$\Delta S_{na}[c] = \Delta S + \Delta S_{ex}[c]$, one can also express this inequality
as a Clausius type inequality of the form
\beq
\left \langle \Delta S \right \rangle \ge - \left \langle \Delta S_{ex} \right \rangle + \left \langle \Delta \T \right \rangle,
\label{secondlaw}
\eeq
which contains as particular cases, the Clausius form of the second law for transitions between equilibrium states and a modified version of the second law for transitions between NESS \cite{Hatano2001_vol86}.

(iii) As noted above, the equality in the first inequality of Eq.~\ref{ineq4} holds in the adiabatic limit for infinitely slow driving. In this limit the r.h.s. of Eq.~\ref{lag} is zero because there is no lag between the distribution $p_T$ and $\pi_T$. The fact that the inequality can be saturated is essential for identifying Eq.~\ref{secondlaw} as a generalization of the second law of thermodynamics.

\section{Fluctuation theorems from consideration of generating functions}

Generating functions provide an alternate way to understand fluctuation relations without considering trajectories explicitly. Let us introduce the generating functions of $(\Delta \T, \Delta S_a)$ and of $(\Delta \T, \Delta S_{na})$, namely
\bea
 g^{(a)}_T(c,\gamma,\epsilon) &=& \left \langle \delta(c-c_T) e^{-\gamma \Delta \T[c] - \epsilon \Delta S_a [c]} \right \rangle, \\
g^{(na)}_T(c,\gamma,\epsilon) &=& \left \langle \delta(c-c_T) e^{-\gamma \Delta \T[c] - \epsilon \Delta S_{na}[c]} \right \rangle.
\eea
These quantities satisfy deformed master equations of the form :
\bea
\partial_t g^{(a)}_t(c,\gamma,\epsilon) &=& \sum_{c'} g^{(a)}_t(c',\gamma,\epsilon) \left( \frac{w_t^{h_t}(c,c')}{\hat w_t^{h_t}(c,c')} \right)^\epsilon L_t^{h_t}(c',c)+ \gamma (\partial_t \ln \pi_t)(c,h_t)g^{(a)}_t(c,\gamma,\epsilon), \nonumber \\
\partial_t g^{(na)}_t(c,\gamma,\epsilon) &=& \sum_{c'} g^{(na)}_t(c',\gamma,\epsilon) \left( \frac{\pi_t(c',h_t) p_t(c)}{\pi_t(c,h_t) p_t(c')} \right)^\epsilon L_t^{h_t}(c',c) \\ && + \left[ \gamma (\partial_t \ln \pi_t)(c,h_t) + \epsilon \partial_t \ln p_t(c)  \right] g^{(na)}_t(c,\gamma,\epsilon).
\eea
We can check that in the special case where $\gamma = -1$ and $\epsilon = 1$ the solutions are
\bea
g^{(a)}_T(c,-1,1) &=& \hat p_T(c) =  \left \langle \delta(c-c_T) e^{-\Delta B_a[c]} \right \rangle,\label{gen_a} \\
g^{(na)}_T(c,-1,1) &=& p_T(c) =  \left \langle \delta(c-c_T) e^{- \Delta A_{na}[c]} \right \rangle
\label{gen_na}
\eea
where $\hat p_t(c)$ is the solution of the master equation with generator $\hat L_t^{h_t}$ as defined in appendix \ref{dual current}.
Note that Eq~\ref{gen_na} can be transformed to remove the boundary term in $\Delta A_{na}[c]$ in the following way
\bea
p_T(c) &=& \left \langle \delta(c-c_T) e^{- \Delta S_{b} - \Y[c]} \right \rangle,  \label{gen1} \\
&=& \sum_{[c]} \P[c|c_0] p_0(c_0) \delta(c-c_T) \left( \frac{p_T(c_T) \pi_0(c_0,h_0)}{ p_0(c_0) \pi_T(c_T,h_T)} \right)  e^{ - \Y[c]}, \\
&=& \frac{p_T(c) }{ \pi_T(c,h_T)} \sum_{[c]} \P[c] \delta(c-c_T) e^{ - \Y[c]},
\eea
so that we finally get the result for the generating function of $\Y[c]$ already obtained in \cite{Verley2011_vol} :
\beq
\pi_T(c,h_T) = \left \langle \delta(c-c_T) e^{-\Y[c]} \right \rangle.
\label{gen2}
\eeq

Through integration over $c$, we immediately obtain from Eq.~\ref{gen1},
the integrated fluctuation theorem
\beq
\l e^{- \Delta A_{na}[c]} \r   = 1,
\label{IFTna}
\eeq
given that $\Delta A_{na}=\Delta S_b + \Y$.
This integrated fluctuation theorem also follows directly from Eq.~\ref{DFT A_nabis}.
Similarly, by integrating over $c$ in Eq.~\ref{gen2},
we have
\beq
\l e^{- \Y[c]} \r   = 1,
\label{IFTa}
\eeq
 which is nothing but the generalized Hatano-Sasa relation given in Eq.~\ref{modified HS}.
Using the Jensen inequality, we recover from these relations, the second law like inequalities of the previous section.

\subsection{Fluctuation theorems for joint probability distributions}
\label{jointFT}
As shown in \cite{Garcia-Garcia2010_vol82,Garcia-Garcia2012_vol2012}, it is possible to derive fluctuation theorems for joint probability distributions of variables which form parts of the total entropy production even when each variable does not satisfy separately a fluctuation theorem. This approach has many advantages as it offers a unifying principle to recover many fluctuation theorems.
It is straightforward to apply this idea to the general case of an observable $\Delta \A$ of the form of Eq.~\ref{def A}. We assume that this observable can be decomposed into a sum of $m$ observables which are anti-symmetric with respect to the combined action of the tilde and of the star involutions:
$\Delta \A=\sum_{i=1}^{m} \Delta \A_i$, with $\Delta \tilde{\A_i}[c^*]=-\Delta \A_i[c]$ for $i=1..m$. We now have
\bea
P(\Delta \A_1[c]=\Delta \A_1,..,\Delta \A_{m}[c] = \Delta \A_{m}) &=& \sum_{[c]} \prod_{i=1}^{m} \delta(\Delta \A_i - \Delta \A_i[c]) \P[c], \\
     &=&  \sum_{[c^*]} \prod_{i=1}^m \delta(\Delta \A_i + \Delta \tilde{\A_i}[c^*] ) \exp \left( \Delta \A \right) \tilde{\P}[c^*], \nonumber \\
&=&  e ^{\Delta \A} \tilde{P}(\Delta \tilde \A_{1}[c] = -\Delta \A_1,.., \Delta \tilde \A_{m}[c] =-\Delta \A_m). \nonumber
\label{DFT joint}
\eea
where the probability $\tilde P$ is defined by
$$ \tilde{P}(\Delta \tilde \A_{1}[c] = \Delta \A_1,.., \Delta \tilde \A_{m}[c] =\Delta \A_m) = \sum_{[c]} \tilde{\P}[c] \prod_{i=1}^{n} \delta(\Delta \A_i - \Delta \tilde{\A_i}[c] ).  $$
Note that if the observables $\Delta \A_i$ do not satisfy the antisymmetry property with respect to the combined action of the tilde and of the star involutions, we still have a weak form of the fluctuation theorem similar to Eq.~\ref{Ptilde2}.

In the particular case of the decomposition of $\Delta S_{tot}$ into $\Delta S_{na}$ and $\Delta S_{a}$ that are anti-symmetric by total reversal, we have
\beq
\ln \frac{ P(\Delta S_{a}[c] =\Delta S_{a}, \Delta S_{na}[c] = \Delta S_{na})}{\bar{P}(\Delta \bar S_{a}[c] = - \Delta S_{a},\Delta \bar S_{na}[c] = - \Delta S_{na})}= \Delta S_{a}+ \Delta S_{na}.
\eeq

We can also apply the same idea on the decomposition $\Delta A_{na}= \Delta S_{na} - \Delta \T $ obtained in Eq.~\ref{nadb decomp}. Since, $\Delta S_{na}$ and $\Delta \T$  are both antisymmetric under the combination of the duality and the total-reversal ($\Delta \hat{\bar{S}}_{na}[\bar c] = - \Delta S_{na}[c]$ and $\Delta \hat{\bar{\T}}[\bar c] = - \Delta \T[c]$), we have
\beq
\ln \frac{P(\Delta \T[c] = \Delta \T , \Delta S_{na}[c]= \Delta S_{na})}{\hat{\bar{P}}( \Delta \hat{ \bar \T}[c] = - \Delta \T , \Delta \hat{ \bar S}_{na}[c] = - \Delta S_{na})} = \Delta S_{na} -  \Delta \T.
\eeq
In the same way, there is a DFT associated with the decomposition  $\Delta B_a= \Delta S_{a} - \Delta \T$ because both $\Delta S_{a}$ and $\Delta \T$ are also antisymmetric under the duality transformation ($\Delta \hat{S}_{a}[ c] = - \Delta S_{a}[c]$ and $\Delta \hat{\T}[c] = - \Delta \T[c]$). Thus, we have
\beq
\ln \frac{P( \Delta \T[c] = \Delta \T , \Delta S_{a}[c] =\Delta S_{a})}{\hat P(\Delta \hat \T[c] = - \Delta  \T , \Delta \hat S_{a}[c] = - \Delta S_{a})} = \Delta S_{a}  - \Delta \T .\label{DFR_Joint_a}
\eeq

\section{Illustrative examples}

In the following, we illustrate using simple analytical models, the fluctuation relations and the modified second law discussed above. There are two main ways to create non-stationary reference distributions. Either these non-stationary distributions can be created due to the choice of initial conditions or due to a driving force. We illustrate both cases with a driven two states model, and we focus particularly on the case of sinusoidal driving. Besides this two state model, we also study a model for a particle in an harmonic potential and obeying Langevin dynamics.

\subsection{Two states model dynamics}

\subsubsection{Non-stationarity from relaxation due to the initial conditions}

We consider a two states model described by the following master equation:
\beq
\partial_t p_t(a) = -w^{h_t}(a,b) p_t(a) + w^{h_t}(b,a) p_t(b),
\label{2statesME}
\eeq
where the jump rate from state $a$ to state $b$ is denoted $w^{h_t}(a,b)$ in the presence of the driving $h_t$, and $w(a,b)$ in the absence of this driving.
We arbitrarily parametrize the rates as
\beq
\label{rates}
w^{h_t}(a,b)=w(a,b)e^{-h_t/2} \qquad \mbox{and} \qquad w^{h_t}(b,a)=w(b,a)e^{h_t/2},
\eeq
where $h_t$ can be thought of as a force which introduces a biais in the transitions rates. Note also that these rates depend on time only through $h_t$. In order to create a non-stationary distribution, we choose the initial probability distribution to be in state $b$, $p_0(b)$, at an arbitrary value different from the steady state value ($p_{st}(a)=w(a,b)/(w(a,b)+w(b,a))$). As a result, even in the absence of driving, the system will relax in time.

\begin{figure}[H]
\begin{center}
\begin{tabular}{cc}
\parbox[h]{11cm}{\includegraphics[width= 11cm]{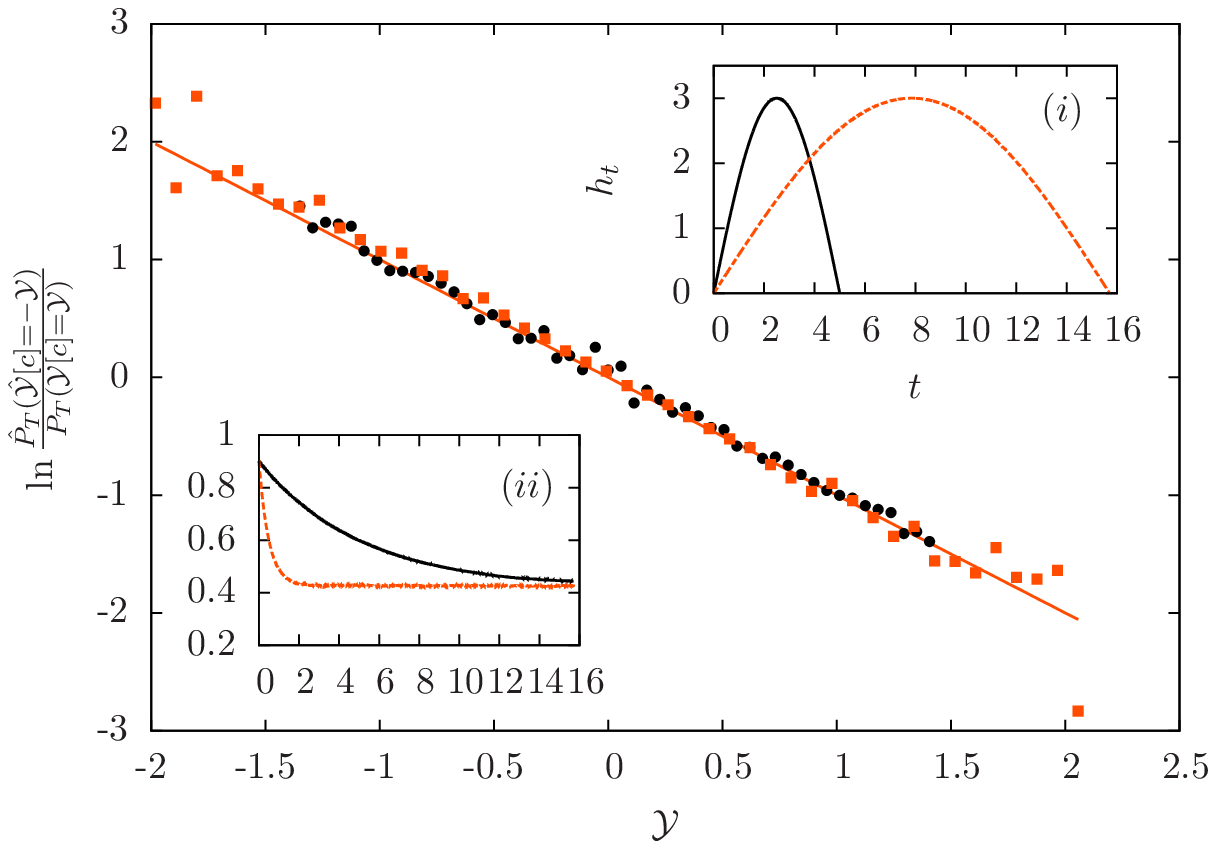}}
\end{tabular}
\end{center}
\caption{Illustration of the detailed fluctuation relation obeyed by the quantity $\Y$. Orange squares correspond to a long protocol with fast relaxation of the $\pi_t$ function towards equilibrium (orange dashed line of the insets) whereas black dots are for a short protocol with slow relaxation towards equilibrium (black solid lines of the insets). Inset $(i)$ shows the half sinusoidal protocols and inset $(ii)$ shows the relaxation, as a function of time $t$, of the distribution $\pi_t(b,h)$ towards equilibrium distribution for a given $h$ value.}
\label{fig1}
\end{figure}

For simplicity, we assume that the driving follows a half sinusoidal protocol depicted in the inset $(i)$ of figure \ref{fig1}, which implies that both the driving and the transition rates are  symmetric with respect to time.
In inset $(ii)$ of figure \ref{fig1}, we show the relaxation of $\pi_t(b,h)$ towards the equilibrium distribution at a given value of $h$ for two choices of the unperturbed rates. One relaxation is faster than the other one because the unperturbed rates are chosen to be larger.

In order to illustrate the DFT of Eq. \ref{DFT A_na}, we have evaluated numerically the functions $\pi_t(c,h)$ for different constant force protocols $h$ with a kinetic Monte Carlo algorithm \cite{Gillespie1977_vol173}. Using this data, we have generated an ensemble $\hat{\bar{\mathcal{C}}}$ of trajectories with the dual reversed dynamics. We have also built separately an ensemble $\mathcal{C}$ of trajectories corresponding to the original dynamics. Then, we have measured the probability of $\Y[c]$ with $[c] \in \mathcal{C}$ and the probability of $\hat{\bar{\Y}}_t[c] = - \Y[\bar c]$ with $[c] \in \hat{\bar{\mathcal{C}}}$ counting the number of times that the value of these functionals were in a given range $ [\Y, \Y+\delta \Y] $. In figure \ref{fig1}, we verify the detailed fluctuation relation for $\Y$ for the slow and the fast protocol of inset $(i)$.
In both cases, the initial condition of the dual reversed experiment was chosen to be $\hat{ \bar p}_T(c) = \pi_T(c,h_T)$ which means that by construction $\Delta S_b=0$ and $\Delta A_{na} [c] = \Y[c] $.
We find that the probability distributions obtained from these simulations follow the expected symmetry. There is one practical difficulty in these simulations, which is also frequently encountered with other numerical tests of fluctuation theorems. One needs to find conditions such that the system is not too far from equilibrium to get a good overlap between $P(\Y)$ and $ \hat{ \bar P}(-\Y)$, and at the same time sufficiently out of equilibrium so that the probability distributions are distinct despite numerical errors.

\subsubsection{Non-stationarity from periodic driving}

\begin{figure}[H]
\begin{center}
\begin{tabular}{cc}
\parbox[h]{11cm}{ \includegraphics[width= 11cm]{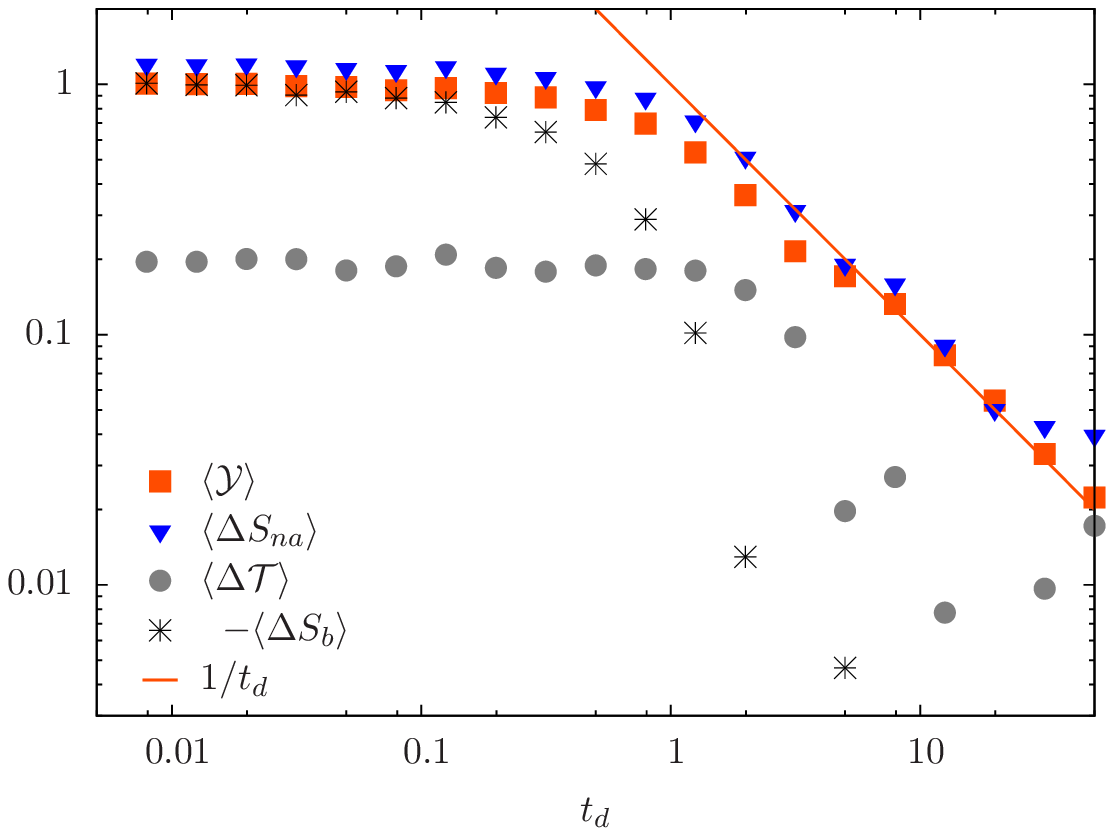}}
\end{tabular}
\end{center}
\caption{Illustration of the modified second law for non-stationary systems created by periodic driving. Symbols represent the driving entropy production $\langle \Y \rangle$ (filled squares), the non adiabatic entropy production $\langle \Delta S_{na} \rangle$ (filled triangles), the difference of traffic $\langle \Delta \T \rangle$ (bullets) and $ \l \Delta S_b \r $ (stars) as function of the duration of the driving $t_d$. The solid line is simply $1/t_d$ and shows that $\langle \Y \rangle$ and $\langle \Delta S_{na} \rangle$ behave as the inverse of the duration of the driving $t_d$ in the limit of large $t_d$.
}
\label{fig2}
\end{figure}

To illustrate the inequalities generalizing the second-law for transitions between non stationary states obtained in Eq.~\ref{ineq4}, we use again the same two states model but now with a different protocol. The shape of the protocols for the driving protocol $h_t$ and the relaxation of $p_t$ towards $\pi_t$ is shown in figure \ref{fig3}: the protocol oscillates around an average value $h^{avg}_t$ which evolves in time following a piecewise protocol of duration $t_d=t_{df}-t_{di}$. This average of the protocol $h^{avg}_t$ represents the real driving which induces a transition from one non stationary state to another one, while the oscillations around the average create these non-stationary states. As before, we have used kinetic Monte Carlo simulations to measure the functions $\pi_t(a,h^{avg})$ for different values of $h^{avg}$. We have then carried out simulations with the time-dependent driving to obtain the quantities $\left \langle \Y \right \rangle $, $\left \langle \Delta \T \right \rangle $ and $\left \langle \Delta S_{na} \right \rangle $ at fixed final time $T$ for different values of the duration of the driving $t_d$. As expected, we observe on figure \ref{fig2} that $\left \langle \Delta S_{na} \right \rangle \ge \left \langle \Delta \T \right \rangle$. When we calculate $ \left \langle \Delta S_b \right \rangle$ at the final time $T$, we find a value close to zero irrespective of the duration of the protocol $t_d$ because the system has either be driven so slowly that $p_t(c,[h])$ has relaxed to $\pi_t(c,h_t)$ already at the end of the protocol at $t_{df}$ or, the system has relaxed afterwards between the times $t_{df}$ and $T$.  This is compatible with $\l \Y \r  \ge - \left \langle \Delta S_b \right \rangle$, but in fact, since $\left \langle \Y \right \rangle$ does not change between the time $t_{df}$ and $T$, one can  obtain a closer bound for $\left \langle \Y \right \rangle$ by evaluating $\left \langle \Delta S_b \right \rangle$ at the final time of the driving $t_{df}$ instead of $T$ as shown in figure \ref{fig2}. For this reason, in figure \ref{fig2}, we show $\langle \Y \rangle$, $\langle \Delta S_{na} \rangle$ and $\langle \Delta \T \rangle$ evaluated between the time $t_{di}$ and time $T$ whereas $ \l \Delta S_b \r $ is evaluated at time $t_{df}$.

In this figure, we also show that the way $\langle \Y \rangle$ approaches the adiabatic limit for large $t_d$ is through a scaling law in $1/t_d$. In fact, this is precisely the scaling found in this limit for the dissipated work for a process starting in equilibrium as function of $t_d$. This dependence has been first theoretically predicted in \cite{Sekimoto1997_vol66}, and recently confirmed in an experiment aimed at testing experimentally the Landauer principle \cite{Berut2012_vol483}.

\begin{figure}[H]
\begin{tabular}{cc}
\parbox[h]{16cm}{ \includegraphics[width= 16cm]{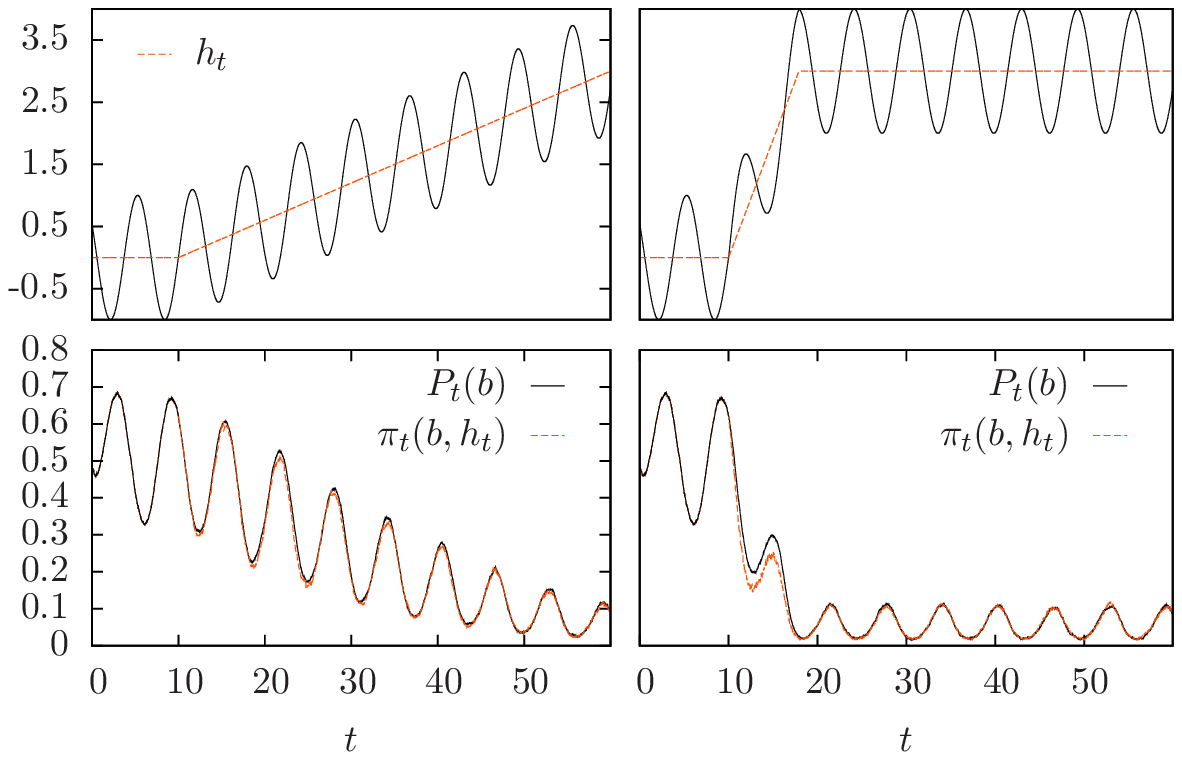}}
\end{tabular}
\label{fig4}
\caption{Transition between non stationary oscillating states for different time of driving: (from left to right) $t_d = 50$ and $t_d=7.91$. Top: Protocol imposed in orange  dashed lines and in black line the total force applied on the system. Bottom: Exact (black) and accompanying probability distribution (orange) of state $b$ as a function of time. For the shortest protocol, we see that the accompanying distribution is different from the exact solution while the protocol is changing.}
\end{figure}

\subsection{Overdamped Langevin dynamics}

In the previous section, we have used the same control parameter to create the non-stationary state and to induce a transition between the initial and the final non stationary states. On the contrary, in this last section, we consider a particle obeying an overdamped Langevin dynamics in an harmonic potential with two different driving forces,  a time-dependent spring constant $k_t$ which oscillates and create a non-stationary periodic state, and a piecewise non conservative time dependent force $h_t$ which acts as driving inducing transitions. The position $x_t$  of the particle is given by the following stochastic differential equation
\beq
\gamma \dot x_t = - k_t x_t + h_t + \eta_t \sqrt{2 \gamma / \beta },
\label{Langevin eq}
\eeq
with $\gamma$ the friction coefficient, $\beta$ the inverse temperature and $\eta_t$ a Gaussian white noise of mean zero and variance unity. In this case, $x_t$ is a gaussian process, which means that the probability distributions $p_t(x,[h_t])$ and $\pi_t(x,h)$ are known from the variance and the average value of the position \cite{Verley2011_vol}. From these quantities, we obtain $\Y$ directly through Eq.~\ref{defY} and the boundary term $\Delta S_b(t_{df})$ from  $\ln \pi_{t_{df}}(x_{t_{df}},h_{t_{df}}) - \ln p_{t_{df}}(x_{t_{df}},[h]) $. As in the other example, we confirm that $\left \langle \Y \right \rangle \ge -\left \langle \Delta S_b \right \rangle$ for all values of the duration of the driving $t_d$.
\begin{figure}
\begin{center}
\begin{tabular}{cc}
\includegraphics[width= 11cm]{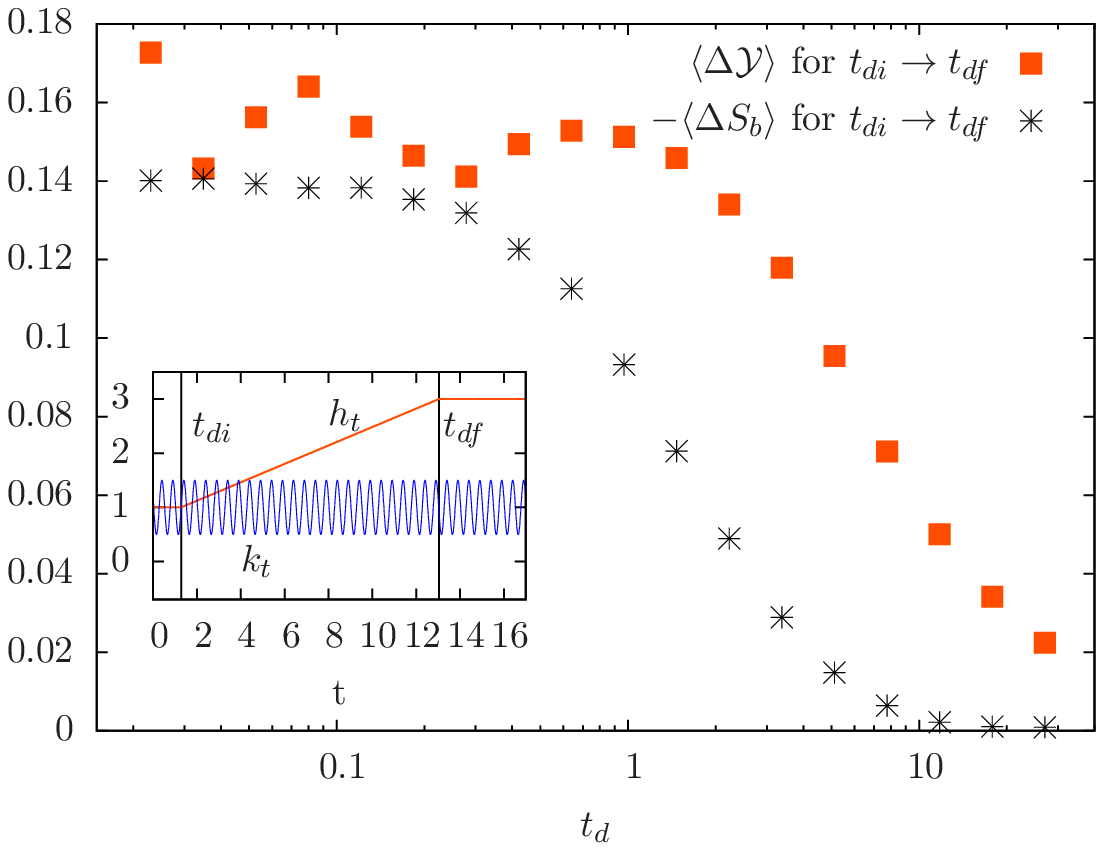}
\end{tabular}
\end{center}
\caption{Inset: Transitions between two non stationary states corresponding to  the forces $h_{t_{di}} = 1.3$ and $h_{t_{df}} = 3$. The spring constant is oscillating with a period of $0.5$ around the value $1$ and with an amplitude of $0.5$. The inverse temperature is taken at $\beta= 0.2$, the friction is $\gamma = 1$.}
\label{fig3}
\end{figure}

In a recent experiment, the heat fluctuations of a brownian particle have been measured in an aging gel, created by a sudden temperature quench \cite{Gomez-Solano2011_vol106}.
This aging gel plays the role of a non-equilibrium bath for the probe particle.
With the same experimental setup, the deviation from the fluctuation-theorem has been measured by evaluating separately the correlations and the response function \cite{Gomez-SolanoJ.R.2012_vol98}.
A complete discussion of these interesting results is out of place here, but instead we show that the detailed fluctuation for the heat exchange obtained in this reference follows from the framework developed in previous sections. The dynamics followed by the probe particule in the experiment is similar to that described by Eq.~\ref{Langevin eq} but differs from it in that in the experiment, there is no driving force and the spring constant is not time dependent. We can adapt the formalism developed in section \ref{sec:path proba} to the experimental situation by choosing a similar logratio of probabilities as in Eq.~\ref{def A} with the tilde operation taken to be the identity, and the star to represent time-reversal. We therefore consider the quantity
\beq
\Delta \rho[x]= \ln \frac{\P[x]}{\P[\bar{x}]}.
\label{def A1}
\eeq
Since the quench, which occurs at time 0, is very fast and there is no subsequent driving in the experiment, the dynamics occurring at time $t>0$ is described by time-independent rates denoted simply $w(c,c')$. If we consider now a path probability ratio with trajectories starting at time $t>0$ and finishing at time $T$, we obtain from Eq.~\ref{explicit A}, that
\beq
\Delta \rho[x] = \ln \frac{p_t(x_t)}{p_t(x_T)}  + \sum_{j \ge j0}^N \ln \frac{w(x_{j-1},x_j)}{w(x_{j},x_{j-1})},
\label{explicit rho}
\eeq
where the index $j_0$ corresponds to time $t$.
Since tilde was chosen to be the identity, it is obvious that the corresponding transformation of the full path probability is an involution. It follows from section \ref{sec:path proba}
that in this case
\beq
\ln \frac{P(\Delta \rho[x]=\Delta \rho)}{P (\Delta \rho[x] = - \Delta \rho)}= \Delta \rho.
\label{DFT rho}
\eeq
Now, the second term in the r.h.s. of Eq.~\ref{explicit rho} corresponds to what is called medium entropy $\Delta S_m$. The temperature of the medium surrounding the probe particle equilibrates very fast (unlike the degrees of freedom associated with the polymers which constitute the gel), so that we can consider that $\Delta S_m=-\beta q$, where $\beta=1/T$, and $T$ is the equilibrium temperature of the surrounding medium, with $q$ is the heat exchanged by the particle and the medium. Since there is no work, this heat is simply the variation of internal energy, so $q=k(x_{t+T}^2-x_t^2)/2$. Furthermore, since the quench is fast and the system was prepared in an equilibrium state before the quench with a gaussian distribution, the distribution of the initial condition at time $t$ is still a gaussian of variance denoted $\sigma_x^2(t)$ in Ref.~\cite{Gomez-Solano2011_vol106}. At time $t+T$, it is assumed that the system is equilibrated so that $k \sigma_x^2(t+T)=k_B T$.
In view of this, we obtain from Eq.~\ref{DFT rho}, the fluctuation relation satisfied by the heat $q$ obtained in this reference with $\Delta \rho=-\Delta \beta q$, and
\beq
\Delta \beta= \frac{k_B T}{k} \left[ \frac{1}{\sigma_x(t+T)^2} -\frac{1}{\sigma_x(t)^2} \right],
\eeq
which is interpreted as an effective temperature imbalance \cite{Gomez-Solano2011_vol106}. This property however only holds for the case of linear Langevin dynamics with a time-independent spring constant.

We thus see on this example that the detailed fluctuation relation satisfied by the heat exchange in Eq.~\ref{DFT rho} follows from general considerations of a logratio of probabilities of the form of Eq.~\ref{def A1}. That this should be the case was also apparent in the derivation of a related fluctuation theorem satisfied by the heat exchange between a system and two thermostats \cite{Jarzynski2004_vol92}.

\section{More complex systems}
When using the theoretical framework developed in this paper for complex systems - such as aging systems -, one will encounter the difficulty already present in the standard Hatano-Sasa that the distribution $\pi_t(c,h)$ (or $p_{stat}(c,h)$ for the standard Hatano-Sasa) is difficult to determine and may not be a smooth function \cite{Perez-Espigares2012_vol85}. Indeed, this distribution can be calculated analytically only in a few simple cases, such as in the case of discrete models involving only a few states or for a particle in an harmonic trap obeying overdamped Langevin dynamics as discussed in the previous section \cite{Gomez-Solano2011_vol106,Gomez-SolanoJ.R.2012_vol98}. For more complex systems, this distribution will not be accessible analytically. However if the system (or sub-system) of interest is of small size, the numerical determination of this distribution is possible through extensive simulations as we have shown on an example based on the Glauber-Ising model \cite{Verley2011_vol}. Among the various other strategies which can facilitate this numerical determination, one recent interesting suggestion is to determine the distribution iteratively by starting from an approximate ansatz function \cite{Perez-Espigares2012_vol85}.

\section{Conclusion}

In this paper, we have first emphasized a particular point, namely that a
 detailed fluctuation theorem can be of strong or weak form depending on whether
 the initial and final probability distributions have a symmetry under protocol reversal.
As we discussed in the case of the entropy production, this property means that the system stochastic entropy is or not anti-symmetric with respect to protocol reversal.

We have then presented a general framework for systems which are prepared in a non-stationary non-equilibrium state in the absence of any perturbation, and which are then further driven through the application of a time-dependent perturbation. Typically for applications, this perturbation is applied as a means to probe the non-equilibrium properties of the unperturbed non-equilibrium state. We can formally distinguish two different situations depending on the way the non-equilibrium state is prepared.

In the first category, the non-equilibrium state is created by some driving, and thus the perturbation which will be applied to it after some time should be viewed as a second driving. As a particular simple example of this category, one can create the initial state by a periodic driving. In these conditions, our approach predicts a modified second law of thermodynamics for transitions between periodically driven states. Such periodically driven states are achievable in a number of experimental systems such as vibrated granular medium, electronic circuits, manipulated colloidal systems, or quantum optics for instance.

In the second category, the initial non-stationary state is a transient state produced by the choice of initial conditions. For instance, the system has been prepared by a quench of some parameter which can be the temperature or the concentration for instance, and the dynamics which follows involves relaxation or coarsening. This is typically what happens in a glassy system, where the slow relaxation following this quench leads to aging.

For all these systems, the generalization of the second law of thermodynamics derived in this paper should hold. In this extension, the dissipated work which enters one form of the standard second law is replaced by the average of a new functional $\langle \Y \rangle$, which can be defined without reference to thermodynamics. We found that this quantity is related to the lag between the actual probability distribution and the $\pi_t$ distribution evaluated at the current value of the control parameter, in the same way as the dissipated work is related to the lag with respect to the equilibrium distribution. Furthermore, $\langle \Y \rangle$ approaches the adiabatic limit in a similar way as the dissipated work, ${i.e.}$ in a manner which is proportional to the inverse of the duration of the driving.
We hope that our work can contribute to the elaboration of a theoretical framework for modified fluctuation-dissipation theorem and modified second law,
in particular for systems in contact with a non-equilibrium bath.

\subsection*{Acknowledgements}
We thank U. Seifert, M. Esposito, C. van den Broeck, R. Ch\'etrite and A. Kundu for many insightful discussions in connection with this work.


\appendix


\section{Definition of duality from current reversal}
\label{dual current}

In the main text, we have introduced four dynamics with generators $L_t^{h_t}$, $\bar L_t^{h_t}$, $\hat L_t^{h_t}$ and $\hat{\bar{L}}_t^{h_t}$. For all these dynamics, with a generator that we write generally $\tilde L_t^{h_t}$ to encompass all cases, we define a
probability distribution $\tilde p_t(c)$  solution of the following master equation :
\beq
\frac{d \tilde p_t(c)}{dt} = \sum_{c'}  \tilde p_t(c')  \tilde L^{h_t}_t(c',c).
\label{defptilde}
\eeq
In the same spirit, we have several reference probability distributions $\pi_t(c,h), \bar \pi_t(c,h), \hat \pi_t(c,h)$ and $\hat{\bar \pi}_t(c,h)$ associated to the generators $ L_t^{h}, \bar L_t^{h}, \hat L_t^{h}$ and $\hat{\bar L}_t^{h}$, that we note generally $\tilde \pi_t(c,h)$ with generator $\tilde L_t^{h}$. The corresponding general master equations is
\beq
\left( \frac{\partial \tilde \pi_t}{\partial t}  \right) (c,h) = \sum_{c'} \tilde \pi_t(c',h) \tilde L_t^{h}(c',c) = \sum_{c'} \tilde \J_t^{h_t}(c,c'),
\label{defpitilde}
\eeq
in which we have defined the reference probability current of the dynamics modified by the tilde transformation
\beq
\tilde{\J}^{h}_t(c,c') = \tilde {\pi}_t(c,h)\tilde{w}^{h}_t(c,c') -\tilde{\pi}_t(c',h)\tilde{w}^{h}_t(c',c).
\eeq
We want to show in this appendix that the dual dynamics corresponds to the dynamics for which accompanying probability currents in the system at time $t$ are opposite to accompanying probability currents in the system with reversed dynamics at time $T-t$, that is to say \cite{Crooks2000_vol61,Chetrite2008_vol282}:
\beq
\hat{ \J}^{h_t}_t(c,c')  = - \bar{\J}^{ h_{T-t}}_{T-t}(c,c').
\label{current reversal}
\eeq
To do so, we start from this definition of duality and find back the dual rates of Eq.~\ref{duality}. First we remark that if all dynamics are connected, it is the same for the reference probability distributions. For instance, we can check that $\hat{\bar \pi}_{\tau}(c,h_\tau)=\pi_{T-\tau}(c,h_{T-\tau})$ by verifying that both quantities are the solution of the same differential equation
\beq
(\partial_t \hat \pi_t ) (c',h_t) = \sum_{c} \hat{ \J}^{h_t}_t(c,c') = - \sum_{c} \bar{ \J}^{h_{T-t}}_{T-t}(c,c') = - \left(\partial_{(T-t)} \bar \pi_{T-t}\right)(c',h_{T-t}) 
\eeq
with the same initial condition $\hat{\bar \pi}_{0}(c,h_0)=\pi_{T}(c,h_T)$. Now, to obtain the dual rates, we use this symmetry $\hat{\pi}_{\tau}(c,h_\tau)= \bar{\pi}_{T-\tau}(c,h_{T-\tau})$ and Eq~\ref{current reversal} to get
\bea
\hat {\pi}_t(c,h_t)\hat{w}^{h_t}_t(c,c') -\hat{\pi}_t(c',h_t)\hat{w}^{h_t}_t(c',c) &=& - \left( \hat {\pi}_t(c,h_t)\bar{w}^{h_{T-t}}_{T-t}(c,c') -\hat{\pi}_t(c',h_t)\bar{w}^{h_{T-t}}_{T-t}(c',c) \right), \nonumber \\
&=& \hat{\pi}_t(c',h_t)w^{h_{t}}_{t}(c',c) - \hat {\pi}_t(c,h_t)w^{h_{t}}_{t}(c,c').
\eea
The simplest rates that verify this equality are
\beq
\hat w^{h}_\tau(c,c')= \frac{w^h_\tau(c',c) \hat \pi_\tau(c',h)}{ \hat \pi_\tau(c,h)}.
\label{dualitybis}
\eeq
The last step consists to use the fact that the duality obtained from Eq.~\ref{current reversal} has to be an involution (whereas it was not trivial to see it on Eq~\ref{duality}) in such a way that $ \hat{ \hat{ w}}^{h}_\tau(c,c') = w^{h}_\tau(c,c')$. We then end with Eq~\ref{duality} as another definition of the duality transformation. Note that we have  $\bar{\J}^{ h_{T-t}}_{T-t}(c,c') \neq  \J^{ h_{t}}_{t}(c,c')$ because $\bar \pi_{t-T}(c,h_{t-T}) \neq \pi_t(c,h)$ as we can check using Eq~\ref{defpitilde} so duality is not a trivial reversal of the current as it was in the stationary reference framework.

\bibliographystyle{unsrt}
\bibliography{Ma_base_de_papier}

%

\end{document}